\documentclass[conference]{IEEEtran}
\usepackage{cite}
\usepackage{amsmath,amssymb,amsfonts}

\usepackage{algorithm}
\usepackage{algorithmic}
\usepackage{bbold}
\usepackage{graphicx}
\usepackage{textcomp}
\usepackage{braket}
\usepackage{xcolor}
\usepackage{lipsum}
\usepackage{epsfig}
\usepackage{tikz}
\usetikzlibrary{shapes.geometric, arrows, quantikz, backgrounds, patterns, shapes.symbols, decorations.pathmorphing}
\def\BibTeX{{\rm B\kern-.05em{\sc i\kern-.025em b}\kern-.08em
    T\kern-.1667em\lower.7ex\hbox{E}\kern-.125emX}}

\begin{document}

\title{Comparative Benchmark of a Quantum Algorithm\\ for the Bin Packing Problem}

\author{\IEEEauthorblockN{Mikel Garcia de Andoin\IEEEauthorrefmark{1}\IEEEauthorrefmark{2}\IEEEauthorrefmark{3}\IEEEauthorrefmark{6}, Izaskun Oregi\IEEEauthorrefmark{2}, Esther Villar-Rodriguez\IEEEauthorrefmark{2}, Eneko Osaba\IEEEauthorrefmark{2} and Mikel Sanz\IEEEauthorrefmark{1}\IEEEauthorrefmark{3}\IEEEauthorrefmark{4}\IEEEauthorrefmark{5}\IEEEauthorrefmark{6}}\IEEEauthorblockA{\IEEEauthorrefmark{1}Department of Physical Chemistry, University of the Basque Country UPV/EHU, 48940 Leioa, Spain\\}\IEEEauthorblockA{\IEEEauthorrefmark{2} TECNALIA, Basque Research and Technology Alliance (BRTA), 48160 Derio, Spain\\}\IEEEauthorblockA{\IEEEauthorrefmark{3}EHU Quantum Center, University of the Basque Country UPV/EHU, 48940 Leioa, Spain\\}\IEEEauthorblockA{\IEEEauthorrefmark{4}IKERBASQUE, Basque Foundation for Science, 48009 Bilbao, Spain\\}\IEEEauthorblockA{\IEEEauthorrefmark{5}Basque Center for Applied Mathematics BCAM, 48009 Bilbao, Spain\\}\IEEEauthorblockA{\IEEEauthorrefmark{6}Corresponding authors: mikel.garciadeandoin@ehu.eus, mikel.sanz@ehu.eus}
}

\maketitle

\begin{abstract}
The Bin Packing Problem (BPP) stands out as a paradigmatic combinatorial optimization problem in logistics. Quantum and hybrid quantum-classical algorithms are expected to show an advantage over their classical counterparts in obtaining approximate solutions for optimization problems. We have recently proposed a hybrid approach to the one dimensional BPP in which a quantum annealing subroutine is employed to sample feasible solutions for single containers. From this reduced search space, a classical optimization subroutine can find the solution to the problem. With the aim of going a step further in the evaluation of our subroutine, in this paper we compare the performance of our procedure with other classical approaches. Concretely we test a random sampling and a random-walk-based heuristic. Employing a benchmark comprising 18 instances, we show that the quantum approach lacks the stagnation behaviour that slows down the classical algorithms. Based on this, we conclude that the quantum strategy can be employed jointly with the random walk to obtain a full sample of feasible solutions in fewer iterations. This work improves our intuition about the benefits of employing the scarce quantum resources to improve the results of a diminishingly efficient classical strategy. 
\end{abstract}


\section{Introduction}

Optimization is an intensively studied research field, which is the main focus of hundreds of papers annually. The main interest of this knowledge area lies on the wide range of real-world applications it can cover. An efficient dealing of this kind of problems usually involves the needing of remarkable computational resources, making brute-force method impractical for even medium-sized instances. As a result of this situation, a plethora of techniques have been proposed along the years for competently address optimization problems. As can be thought, almost plenty of optimization solvers are conceived for their execution on classical computation systems. However, Quantum Computing (QC, \cite{steane1998quantum}) is progressively emerging as a promising alternative to this classical devices, providing a revolutionary approach for tackling complex optimization problems.

Two tasks where, QC has demonstrated its potential are factorization and unstructured search problems. With the aim of factorizing large integers, a NP-hard problem that takes exponential time to be solved, Shor proposed a quantum algorithm (Shor's algorithm \cite{Shor}) capable of solving it in polynomial time. Later in 1996, Grover presented a quantum search framework that finds a particular data in an unstructured database using fewer evaluations than its classical counterpart \cite{Grover}. Specifically, Grover’s algorithm was proven to provide a quadratic speedup over the optimal classical algorithm. However, there are still no quantum algorithms that are proven to provide a speedup for optimization. Even though there exist algorithms such as the variational quantum eigensolver (VQE \cite{peruzzo2014variational}) or the quantum approximate optimization algorithm (QAOA \cite{Farhi2014QAOAEXTRA}) that have already shown promising results, there is still room for the exploration of new approaches.

Although the goal of QC is to search for the quantum advantage, the quantum machines that are available today are not error-corrected yet. Commonly denoted as noisy intermediate-scale quantum (NISQ) devices, current machines are composed of tens to hundreds of noisy qubits that perform imperfect operations in a limited coherence time. These limitations make difficult the efficient resolution of real-world problems. Pending scientific progress in developing fault-tolerant devices, one of the main goals of the present NISQ era is to design new frameworks that exploit the power of current quantum devices to solve challenging tasks such as combinatorial optimization problems \cite{bharti2021noisy}.


The experimentation proposed in this paper is focused on the Bin Packing Problem (BPP, \cite{garey1981approximation}). Being one of the most well-known combinatorial optimization problems, the BPP is still employed in a wide range of industrial applications. More specifically, we deal in this work with the one-dimmensional BPP (1dBPP, \cite{munien2020metaheuristic}). 

In this paper we study the advantages of using a quantum algorithm for sampling feasible partial solutions in the hybrid algorithm we previously developed addressing the 1dBPP \cite{mikelgda2022BPPEXTRA}. In short, this strategy consists of using a subroutine to reduce the search space for a second subroutine building the solutions to the problem. This space reduction strategy employs a quantum annealing algorithm to obtain configurations for single containers, which are indeed the partial solutions for the problem. Although we showed that this algorithm was valid for solving the 1dBPP, we did not show how it improves any other classical algorithm. To answer this open question, we implemented two classical sampling algorithms: a random sampling strategy, and a heuristic based on a random walk. We also propose a hybrid sampling strategy combining the random walk and the quantum annealing, which might improve the joint performance and resource consumption. Supported by the results of a benchmark consisting on two runs over 18 instances of size 10 and 12 packages, we show that employing the quantum annealing strategy can provide a full sampling of the feasible partial solutions in less runs compared to the classical ones. Furthermore, we show that the probability of obtaining new feasible for the quantum annealing is close to constant as the sampling is performed. This offers a possible advantage over the random walk, which shows an expected stagnation behaviour for sampling the last feasible partial solutions. 

The rest of the paper is organized as it follows. In section \ref{sec:PreviousWork}, we review the full hybrid algorithm for the 1dBPP, and in particular the quantum annealing subroutine for the sampling. Section \ref{sec:Sampling} defines the problem of sampling feasible partial solutions, in particular for the 1dBPP and in general for variations of the problem. In section \ref{sec:algSamplingSubroutine}, we detail the different classical algorithms we propose for the sampling subroutine. Then, section \ref{sec:results} discusses the results obtained by performing the sampling on a benchmark. Finally, we conclude this article by summarizing the work done, and proposing possible future research.   

\section{Previous work}\label{sec:PreviousWork}

As mentioned before, the problem addressed in this paper is the BPP. In a nutshell, in the BPP, we have a set of packages with certain dimensions which must be shipped into containers of equal size. The objective is to assign these packages to the minimum amount of containers without overflowing any of them. Despite several complex variants of the BPP can be found in the literature, such as the three dimensional one, we tackle in this paper the canonical one-dimensional version, the 1dBPP \cite{munien20201dbpp}. In this case, the packages $i\in\{1,\dots,n\}$ are defined by their weights $\{w_i\}$, and the containers by their maximum capacity $C$. Let's use a simple codification for the solutions, where we encode the package $i$ being assigned to the container $j\in\mathbb{N}$ with $x_i^{(j)}=1$, and $x_i^{(j)}=0$ otherwise. This way, we define 1dBPP as the problem with the objective function
\begin{equation}\label{eq:costfunc}
    \min{b}
\end{equation}
subject to
\begin{align}
    &\sum_{i=1}^{n} w_{i}  x_i^{(j)} \leq C,\ \forall j \in \{1, \dots, b\},\label{eq:const1}\\
    &\sum_{j=1}^{b} x_{i}^{(j)} = 1,\ \forall i  \in \{1, \dots, n\},\label{eq:const2}\\
    &x_i^{(j)} \in \{0,1\},\ \forall i\in \{1, \dots, n\},\ \forall j\in\{1,\dots,b\},\label{eq:const3} 
\end{align}
where $b$ is the total number of containers used. The constraint in Eq. \eqref{eq:const1} refers to the maximum capacity of the containers, Eq. \eqref{eq:const2} imposes that all packages are assigned to a container, and Eq. \eqref{eq:const3} is forced due to the codification we have employed. 
    
For solving the 1dBPP, we employ in this paper our previously published \textit{hybrid quantum-classical algorithm}, which is divided into two main subroutines \cite{mikelgda2022BPPEXTRA}. The first step of this method is to search for the set of configurations for a single container, which we call feasible partial solutions. As we just want the configuration for single containers, we only need to enforce the first constraint (Eq. \ref{eq:const1}). The second phase of the algorithm takes this set of partial solutions and combines them in order to build solutions to the full problem. In this step, we both employ the cost function (Eq. \ref{eq:const1}) and enforce the solutions to fulfill the second constraint (Eq. \ref{eq:const3}). This second part of the algorithm is performed by a classical subroutine. 

\begin{figure}
    \centering
    \resizebox{\linewidth}{!}{
    \begin{tikzpicture}[node distance=1.5cm]    
        \node[ellipse, draw, blue, pattern = north west lines, pattern color=blue, minimum width = 30pt, minimum height = 15pt] (elipse) at (0,0) {};
        \node[cloud, draw, minimum width = 80pt, minimum height = 50pt, cloud puffs = 28] (solution) at (0,0) {};
        
        \draw[line width=1pt, magenta] (0.2,0) -- (0.2,0.2);
        \draw[line width=1pt, magenta] (0.1,0.1) -- (0.3,0.1);
        \draw[line width=1pt, magenta] (-0.1,-0.2) -- (-0.1,0);
        \draw[line width=1pt, magenta] (-0.2,-0.1) -- (0,-0.1);
        
        \node[cloud, draw, minimum width = 80pt, minimum height = 50pt, cloud puffs = 28, aspect = 1.33] (full) at (-4,6) {};
        \node[] (c) at (-4,6) {Full solutions $\mathcal{C}$};
        
        \node[rounded corners, draw, minimum width = 80pt, minimum height = 50pt] (partial) at (4, 6) {};
        \node[] at (4, 6) {Partial solutions $\tilde{\mathcal{C}}$};
        
        \node[rounded corners, draw, minimum width = 80pt, minimum height = 50pt] (feasible) at (2, 3) {};
        \node[diamond, draw, red, pattern = north west lines, pattern color=red, minimum width = 40pt, minimum height = 15pt] (diamante) at (2, 3) {};
        
        \draw[double,thick,-{Triangle[scale=0.6]}] (full) -- node [midway, left = 3pt]{Classical optimization} (elipse);
        
        \draw[thick, decorate, decoration={snake, amplitude=1pt, segment length=3pt,post length=2pt},-{Triangle[scale=0.6]}] (partial) -- node [pos = 0.35, right = 3pt]{Quantum sampling} (diamante);
        
        \draw[double,thick,-{Triangle[scale=0.6]}] (diamante) -- node [pos = 0.6, right = 3pt]{\shortstack{Full solution generation\\ {\footnotesize (Composition of partial solutions)}}} (elipse);
        
        \node[red] (tagFeasible) at (5,3) {\shortstack{Feasible\\partial\\solutions\\$\tilde{\mathcal{F}}$}};
        \draw[-latex, red] (diamante) -- (tagFeasible);
        
        \node[blue] (tagFull) at (4,0) {Feasible full solutions $\mathcal{F}$};
        \draw[-latex, blue] (elipse) -- (tagFull);
        
        \node[magenta] (tagOptimal) at (-4,0) {Optimal solutions $\mathcal{S}$};
        \draw[-latex, magenta] (0.2,0.1) -- (tagOptimal);
        \draw[-latex, magenta] (-0.1,-0.1) -- (tagOptimal);  
    \end{tikzpicture}}
    \caption{Diagram showing the differences between a usual classical approach and our hybrid approach. In many classical optimizers, the algorithm search the solution over the space of full solutions. In our approach, we employ a quantum (or hybrid quantum-classical) algorithm to sample on the set of feasible partial solutions. Then, we use a classical optimizer working on a reduced search space which generates the solutions to the full problem.}
    \label{fig:diagram}
\end{figure}
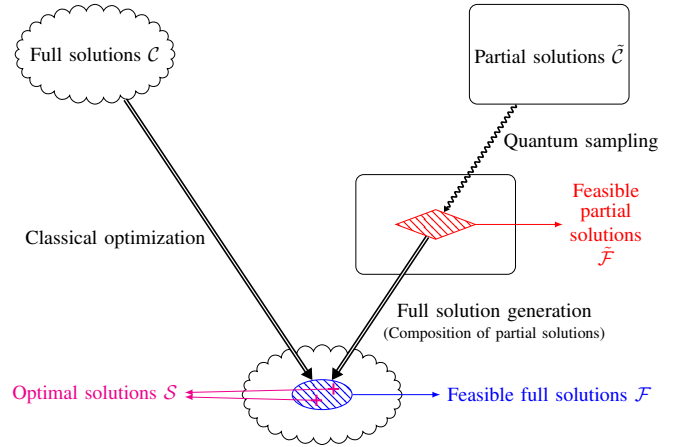

The main idea behind this two-phase strategy is to reduce the initial search space to be used by the classical algorithm from a exponentially large space to a reduced search space (Fig. \ref{fig:diagram}). If we can implement the first subroutine efficiently, then we could remove the computational bottleneck of generating feasible solutions. On this regard, our objective is to delegate the task of obtaining the feasible partial solutions to quantum subroutines, increasing the performance of the algorithm.

\subsection*{Quantum annealing subroutine}
For the problem of subset sampling, we proposed a quantum annealing algorithm. As the objective of this sampling is to find configurations that fulfils the first constraint, its encoding requires less qubits than the full problem, which is a requirement for implementing the algorithm in NISQ devices. 

Usually, inequality constraints can be encoded into a Hamiltonian by employing slack variables \cite[Sec. 4.1.3]{ConvexOptimization}. This converts the problem into an unconstrained problem, with the expense of introducing extra variables to the system. For our annealing algorithm, we employed a simple quadratic Hamiltonian where we only encoded the package weights. If a package is added to the container, then the energy of the system increases with the weight of the corresponding package. Then, we can select an objective weight $T$, so that the sum of the weights of the packages is set in the range of allowed values $0<T<C$. This way, the ground state of our Hamiltonian would encode the possible configurations of packages which fulfil $\sum_iw_ix_i=T$. For this, the problem Hamiltonian we designed is
\begin{equation}\label{eq:problemHamiltonian}
\begin{split}
    H_\text{P}'&=\alpha\left(\sum_iw_ix_i-C\right)+\beta\left(\sum_iw_ix_i-C\right)^2\\
    &=\sum_{i<j}\frac{\beta w_{i}w_{j}}{2}\sigma_i^z\sigma_j^z-\sum_iw_i\left(\frac{\alpha}{2}+\beta\epsilon_w\right)\sigma_i^z,  
\end{split}
\end{equation}
where $x_i\equiv(\mathbb{1}_i-\sigma_i^z)/2$, $\epsilon_w\equiv\sum_iw_i/2-C$, and $\sigma^z_i$ is the Pauli z operator acting on qubit $i$. The objective weight is controlled by both of the parameters $\alpha$ and $\beta$, such that $T=\sum_iw_ix_i=C-\alpha/2\beta$. If we fix $\alpha$ and $\beta$ we would measure the same states in successive runs. Thus, to have access to all feasible solutions, we fix the parameter $\beta$ and we performed a sweep over $\alpha\in[0,2\beta C]$. Furthermore, to avoid measuring the same state twice, we penalize the states that already have been measured. This can be ideally achieved by a projection Hamiltonian
\begin{equation}\label{eq:implementHamiltonian}
    H_\text{P}=H_\text{P}'-\gamma\prod_{\Psi_m\in\mathcal{M}}\ket{\Psi_m}\bra{\Psi_m},
\end{equation}
where $\mathcal{M}$ are the set of measured states, $\Psi_m$ are in the computational basis, and $\gamma$ is the penalty factor.

Now that we have the solution to our sampling problem encoded into the ground state of a Hamiltonian, we need a way to access it. For this, we employ a quantum annealing algorithm, which consists of adiabaticaly changing the Hamiltonian of the system from an initial Hamiltonian $H_0$ to the problem Hamiltonian $H_\text{P}$. If we perform this change infinitely slow, the adiabatic theorem \cite{Born1928Tadiabatico, Kato1950Tadiabatico} assures that the system will remain in its instantaneous eigenstate. In particular, if we start in the ground state of $H_0$, the final state of the system after an adiabatic process would be the ground state of $H_\text{P}$, which we can then measure at the end of the circuit. Even if we can not let the system evolve for an infinite time, we can perform this process in a finite time $T$
\begin{equation}
    U=\exp \frac{-i}{\hbar}\mathcal{T}\int_0^Tdt\left[\left(1-\lambda(t)\right)H_0+\lambda(t)H_P\right],
\end{equation}
where the mixing function $\lambda(t)\in[0,1]$ is strictly increasing. Employing a finite time introduces an error in the fidelity of $F=1-\epsilon(T^{-1})$ \cite{ConsistencyAdiabatic}. For this, we have to select an annealing time that satisfies
\begin{equation}
\label{eq:2.2-AbiabaticCondition}
    T\gg\frac{|\dot{H}(t)|_{max}}{\Delta E^2_{min}},
\end{equation}
where
\begin{equation}
\label{eq:2.2-AdiabaticCondDerivative}
    |\dot{H}(t)|_{max}=\max_{t}\left|\bra{\Psi_0(t)}\frac{dH(t)}{d(t/T)}\ket{\Psi_1(t)}\right|,
\end{equation}
and
\begin{equation}
    \Delta E^2_{min}=\min_{t}(E_1(t)-E_0(t))^2,
\end{equation}
where $\Psi_0(t)$ and $\Psi_1(t)$ are the instantaneous ground and first excited states of $H(t)$ respectively, and $E_0(t)$ and $E_1(t)$ their energies.

In a typical quantum annealing process, we let the system continuously change from the initial to the final Hamiltonian. However, we might not have access to a device where we can continuously control its parameters to implement such process. Then, we can aim at simulating it in a gate based quantum hardware, where we have access to a universal set of discrete quantum gates \cite[Sec. 4.5]{NielsenChuang2010}. For that, we first write the digitized version of the continuous annealing process \cite{Barends2016DigitizedAnnealing}. By using the Suzuki-Trotter formula \cite{Trotter1959Trotter, Suzuki1976Trotter}, we can approximate the evolution of a time-dependent Hamiltonian with a set of $n_T$ evolution steps under time-independent Hamiltonians. This step is conceptually similar to the discretization of a integral when obtaining an approximation by means of numerical methods. Similarly as in numerical integration, the discretization error goes to zero in the limit of dividing the process into infinitely many Trotter steps,
\begin{equation}
    U\approx\mathcal{T}\prod_{k=1}^{n_T-1}\exp\frac{-i\Delta t}{\hbar}\left[\left(1-\lambda(k\Delta t)\right)H_0+\lambda(k\Delta t)H_P\right],
\end{equation}
where $\mathcal{T}$ is the time ordering operator, and $\Delta t=T/n_T$.

An additional step to have a fully implementable algorithm is to apply once again the Suzuki-Trotter formula to each of the steps and separate the evolution under the initial and the final Hamiltonian. This way, we obtain a discretized evolution that is implementable with single- and two-qubit gates. The process that is finally implemented in the quantum hardware is
\begin{equation}
    U\approx\mathcal{T}\prod_{k=1}^{n_T-1}e^{\frac{-i\Delta t}{2\hbar}\lambda'(k\Delta t)H_0}e^{\frac{-i\Delta t}{\hbar}\lambda(k\Delta t)H_P}e^{\frac{-i\Delta t}{2\hbar}\lambda'(k\Delta t)H_0},
\end{equation}
where $H_0=\sum_i\sigma_i^x$ is the initial Hamiltonian. The selection of this Hamiltonian is customary since it is easy to prepare a system in its ground state, for example, applying a Hadamard gate to each qubit which has been initialized at the $\ket{0}$ state. The mixing function $\lambda(t)$ can be optimized to maximize the fidelity of the annealing process, but in this work we employed a linear function $\lambda(t)=t/T$. As is the case with classical heuristics, correctly tuning all the parameters of the problem to maximize the fidelity is a difficult task that requires and in depth analysis of the specific problem instance we are addressing. 

\section{Problem of sampling feasible partial solutions}\label{sec:Sampling}

In this section we will go through the details of the feasible partial solution sampling and its intricacies. For defining the problem of sampling, first we have to define the concept of partial solutions, and global and partial constraints. For a better understanding, let us use the BPP as a starting point. A partial solution $\tilde{x}$ is the configuration of packages inside a single container, $\tilde{x}\in\{1,\dots,n\}$. In this case, the only constraint that restricts single container configurations is the one regarding its maximum capacity (Eq. \ref{eq:const1}). Moreover, a full solution to the problem can be built by concatenating partial solutions, $x=\tilde{x}^{(1)}\oplus\tilde{x}^{(2)}\oplus\cdots\oplus\tilde{x}^{(b)}$. At this step, and since all partial solutions already fulfils Eq. \ref{eq:const1}, we can find feasible (including the optimal) solutions by just imposing the second constraint (Eq. \ref{eq:const2}). 

Let us define the sampling problem in an abstract way. For constrained optimization problems, not all possible inputs to the objective function are allowed as a solution. This issue defines a subset of feasible solutions $\mathcal{F}$, which is contained in the space of all possible input configurations $\mathcal{C}$, $\mathcal{F}\subset\mathcal{C}$. The set of optimal solutions to the problem $\mathcal{S}$ can be found inside $\mathcal{F}$, such that $\mathcal{S}\subseteq\mathcal{F}\subset\mathcal{C}$. A similar relation can be found for partial solutions. The set of partial solutions $\tilde{\mathcal{S}}$ from which we can build the full solutions is found inside the set of feasible partial solutions $\tilde{\mathcal{F}}$, $\tilde{\mathcal{S}}\subseteq\tilde{\mathcal{F}}\subset\tilde{\mathcal{C}}$.

When we employ our hybrid algorithm to solve a problem, the first subroutine ideally aims at sampling $\tilde{\mathcal{S}}$. However, as this subroutine has no access to the information about the cost function, the only way to ensure that we can build an optimal solution to the problem is to completely sample $\tilde{\mathcal{F}}$. We argue that performing this sampling has the hardness of a NP-complete problem. It is immediate to see that completely sampling $\tilde{\mathcal{F}}$ is equivalent of a n-SAT problem \cite{Impagliazzo2001kSAT}. As a sketch of a proof, we can study the problem of finding the last partial solution. In this case, the satisfiability problem would consist of the clauses from the original constraints, plus up to $\lvert\tilde{\mathcal{F}}\rvert-1$ clauses with $n$ literals which prevents the solution to be one of the already sampled solutions. Since the n-SAT problem is NP-complete \cite{Cook1971NPCompleteSAT}, we conclude that the problem of sampling is also NP-complete. 

However, the hardness of the sampling increases as we obtain more samples from $\tilde{\mathcal{F}}$. The initial samples of the problem can be easily obtained, either because the size of $\tilde{\mathcal{F}}$ is comparable to the size of $\tilde{\mathcal{C}}$, or because the constraints of the problem allow us to generate trivial solutions. In 1dBPP, trivial partial solutions can be generated by just shipping one package into a container. 

Since we have this huge hardness difference between various situations, we distinguish three kinds of situations:
\begin{itemize}
    \item \textbf{Initial sampling of lightly constrained problems:} Problems in which the constraints allows us to find trivial partial solutions to the problem. We can also include problems where feasible partial solutions can be found after a small number of steps, $\mathcal{O}(\text{poly}(n))$. We label these kind of problems as lightly constrained problems. Indeed, the 1dBPP as defined in section \ref{sec:PreviousWork} falls in this category.
    \item \textbf{Final sampling of lightly constrained problems:} When completing $\tilde{\mathcal{F}}$ for lightly constrained problems, the hardness of the problem scales up to the hardness of n-SAT problems. The increase of the complexity comes from the reduction of the search space, which effectively increases the number of constraints as the sampling is performed, specifically by one n-local constraint per sample.
    \item \textbf{Heavily constrained problems:} Optimization problems can have constraints that can drastically reduce the number of existing feasible partial solutions. This makes generating feasible partial solutions computationally hard. As there is no limit on how restrictive a problem can be, we can have a set of partial constraints which displays the same characteristics as satisfiability problems. For illustrative purposes, we can think of variations of BPP with additional restrictions, such as packages with more dimensions, or requirements on how to arrange the packages. For example, we can have restrictions that forbids two or more packages to be included on the same container, that forbids including more than a certain number of packages from a subset at the same time, or that imposes how the packages must be arranged inside the container.
\end{itemize}

A straightforward conclusion we can extract from this section is that, regardless of the constraints of the problem, obtaining the set of all feasible partial solutions is a hard problem. Exactly performing this process requires a brute-force strategy with cost $\mathcal{O}(2^n)$, or relaxing the requirements and employing a heuristic strategy. Employing this last option prevents us from ensuring that we have fully sampled $\tilde{\mathcal{F}}$, although sampling a fraction of this set might be enough for obtaining a solution with sufficient certainty. 

\section{Algorithms for the sampling problem}\label{sec:algSamplingSubroutine}

An efficient algorithm requires for all its steps to be optimized. In this work, we explore different algorithms to improve the subset sampling subroutine. 

A key characteristic of the sampling subroutine is that it only has information about the partial constraints. This fact prevent us from employing the cost function in any metric to measure the efficiency of the sampling. To overcome this problem, we propose a new metric in which the objective is to fully sample the subset of feasible partial solutions $\tilde{\mathcal{F}}$ in the least amount of calls to an algorithm. Although this is impractical for solving a problem, this metric encapsulates the information about the speed at which the subroutine can obtain new samples. With this new metric, we explore performance of the following algorithms.

\subsection*{Random Sampling}

The most naive strategy for the sampling consists of using a uniform random number generator to obtain random bit strings. If the bit string corresponds to a feasible solution and it has not been obtained yet, then we can add it to the set of samples. This strategy is only assured to converge to the complete subspace $\tilde{\mathcal{F}}$ after an infinite number of runs. Indeed, the probability of measuring a new feasible partial solution in a run decreases with the number of such solutions we have already obtained $b$, such that
\begin{equation}
    P_\text{random}(\lvert\tilde{\mathcal{F}}\rvert,b)=\frac{\lvert\mathcal{F}\rvert-b}{2^n},
\end{equation}
where $n$ is the number the packages of the problem. Then, the probability of completely measuring $\tilde{\mathcal{F}}$ in $M$ runs of the random sampling strategy is 
\begin{equation}
    P_\text{random}(M)\geq\frac{(2^n-\lvert\tilde{\mathcal{F}}\rvert)^{M-\lvert\tilde{\mathcal{F}}\rvert}\lvert\tilde{\mathcal{F}}\rvert!}{2^{nM}},
\end{equation}
which is valid for $M\geq\lvert\tilde{\mathcal{F}}\rvert$.

\subsection*{Classical simulation of the quantum annealing}

An additional classical algorithm that we can also employ to perform the sampling is the classical simulation of the quantum annealing process, as described in Section \ref{sec:PreviousWork}. Although it seems like a naive solution, this algorithm preserves all the properties of the quantum annealing. Indeed, the classical simulation of the algorithm allows the running of the algorithm in a noiseless environment, increasing the fidelity of the process. Thus, it is expected that the performance of the quantum annealing is bounded from above by its simulation. However, the advantage we obtain in fidelity comes at the expense of the exponential amount of computational resources needed, both in memory and in time, $\mathcal{O}(2^n)$.

\subsection*{Random Walk}

A usual strategy for searching in a binary search space is the random walk \cite{Xia2019RandomWalkReview}. In this heuristic, the possible combinations of bit strings are mapped to the vertex of a graph. At each step of the walk, the state of the system jumps to another vertex, randomly chosen from the set of connected vertices. These transitions are defined through a transition matrix, according to a set of rules. The walk stops once we have found an objective vertex. The number of steps to reach an objective (i.e. the hitting time) depends on both the transition matrix, the initial point and the number of objective points \cite{Levene2002RWhittime, Zhongzhi2012RWhittime}.

For solving the sampling of the 1dBPP, we propose an heuristic based on the random walk. As an overview, the heuristic consist of selecting the packages one by one until we complete a container. The algorithm starts by selecting one of the packages and adding it to the container. In the following steps, the algorithm updates the set of packages that can be added to the container and not overflow it. Then it makes a choice between adding one of the packages or stopping the sample. A feature of this sampling heuristic is that every configuration it yields is a feasible partial solutions. However, a disadvantage of this heuristic is that we can not prevent the algorithm outputting already obtained partial solutions. This decreases the probability of measuring new partial solutions as we perform the sampling. We show in Algorithm \ref{alg:random} the pseudo-code of this algorithm for a single run. 

\begin{algorithm}
\caption{Random walk heuristic subroutine iteration}\label{alg:random}
\begin{algorithmic}
\REQUIRE Package sizes $\{w_i\}$, container size $C$, sampled feasible partial solutions $\mathcal{P}$
\STATE Select one of the packages uniformly at random, $\text{\texttt{candidate}}\in\{1,\dots,n\}$
\STATE Initialize set of eligible packages, $\text{\texttt{left}}=\{1,\dots,n\}/\text{\texttt{candidate}}$
\WHILE{$\text{\texttt{left}}\neq\emptyset$}
    \STATE Calculate current weight, $\text{weight}=\sum_{i\in\texttt{candidate}}w_i$
    \STATE \# \textit{Delete the packages that would overload the container if we add them.}
    \FOR{$i\in\text{\texttt{left}}$}
        \IF{$\text{weight}+w_i>C$}
            \STATE Remove package $i$ from \texttt{left}
        \ENDIF
    \ENDFOR
    \STATE \# \textit{End the walk or select a new package}
    \IF{$\text{random number}\in[0,1]<1/(\text{size of \texttt{left}}+1)$}
        \STATE Break while loop
    \ELSE
        \STATE Randomly pick one package from \texttt{left} and add it to \texttt{candidate}
    \ENDIF
\ENDWHILE
\IF{\texttt{candidate} not in $\mathcal{P}$}
    \STATE Add \texttt{candidate} to the solutions list, $\mathcal{P}=\mathcal{P}\cup\text{\texttt{candidate}}$ 
\ENDIF
\ENSURE Updated list of all sampled subsets, $\mathcal{P}$
\end{algorithmic}
\end{algorithm}

One key point of the proposed random walk heuristic is that we have access to trivial partial solutions, i.e. shipping a single package in a container. From this trivial solution, we have generated a strategy on a graph which guarantees that all feasible partial solutions are connected. If we had a heavily constrained variation of the 1dBPP, we might not be able to generate a graph for the random walk with such characteristics. As we will show in section \ref{sec:results}, for the case of lightly constrained problems, this heuristic is efficient at the start of the sampling process. However, as we obtain more feasible partial solutions, the probability of obtaining a new result decreases drastically.

\subsection*{Hybrid random walk-quantum annealing}

As we have previously mentioned, the random walk strategy is efficient only for the initial sampling of lightly constrained problems. With regard to the quantum annealing algorithm, assuming a perfect adiabatic process with an ideal (and unachievable in practice) hardware, we would obtain a new feasible solution in each run. However, employing quantum devices to solve problems that are easily solvable classically might be a waste of resources. 

Having these two ideas in mind, we propose to combine both algorithms to reduce the total number of iterations we need to have a quality output for the subroutine. Thus, for the first part of the sampling, we employ the random walk. However, there would be a point of the subroutine in which obtaining a new partial solution suppose a significant increase in the number of iterations. At this moment, the proposed hybrid method switches to the quantum annealing algorithm. In order to maximize the probability of measuring new results with the annealing strategy, we keep track of every solution obtained in the random walk. In the first iteration of the annealing, instead of starting only with the bare Hamiltonian from equation \ref{eq:problemHamiltonian}, we add the corresponding penalty terms from the solutions obtained from the random walk (Eq. \ref{eq:implementHamiltonian}).

For defining the switching point, we will assume that the random walk stops being efficient when the estimate of the probability of measuring new results is lower than a certain threshold or when its behaviour deviates from being linear. We quantify this last condition with the root-mean-square deviation (RMSD) from a linear fit. As the random walk is subject to random fluctuations, and since we expect the random walk to be efficient at the start of the sample, we also set a minimum amount of iterations we let it run until the switching point.

\section{Results}\label{sec:results}

\begin{figure}
    \centering
    \includegraphics[width=\linewidth,trim={0pt 0pt 0pt 0pt},clip]{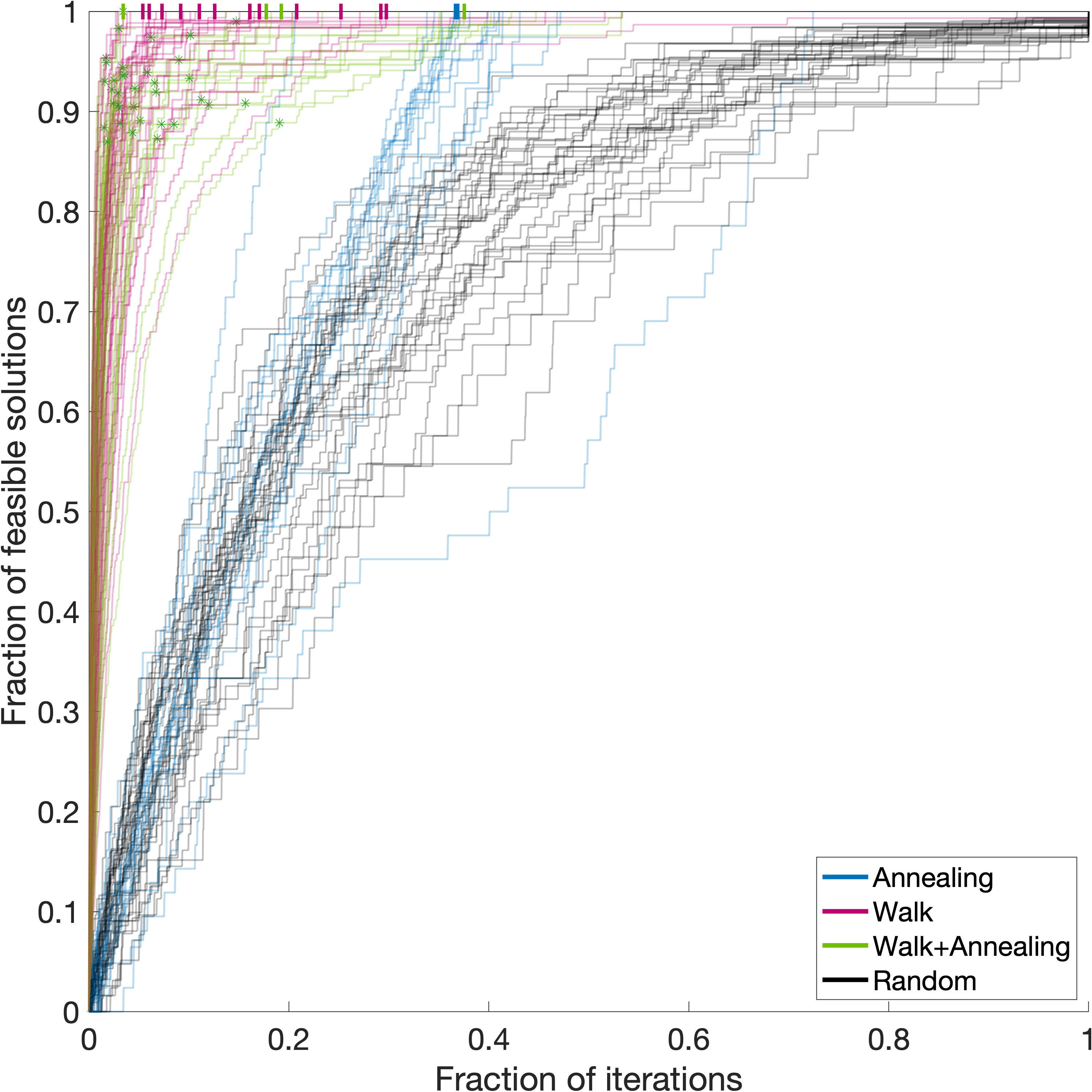} 
    \phantom{a}
    \includegraphics[width=\linewidth,trim={0pt 0pt 0pt 0pt},clip]{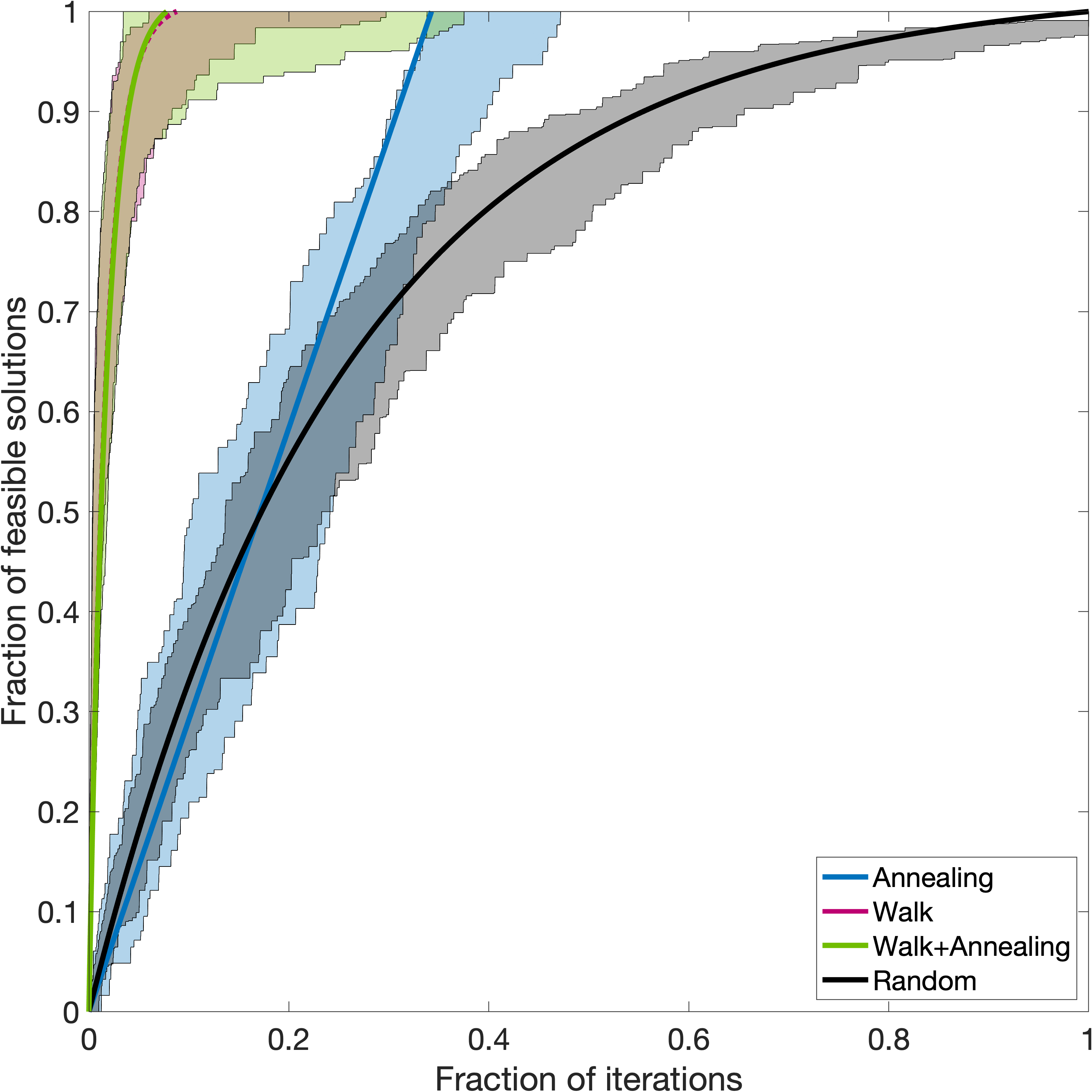}
    \caption{Results of the benchmark over 18 instances of of 1dBPP with 10 and 12 packages. Here we plot the fraction of different feasible partial solutions versus the fraction of iterations for each algorithm and for each instance, calculated as \texttt{algorithm iterations/max iterations to solution}. In the top plot, each line represents one of the two runs for each instance of the different sampling strategies (see legend). The small crosses marks the point at which the hybrid algorithm switched from the classical walk to the quantum annealing. The vertical lines in the upper axis represent the fraction of iterations in which the fastest algorithm obtained the solution for each instance for the problems of length 10 (the classical simulation of the quantum annealing is too slow to reach the solution in a reasonable time). In the bottom plot, we represent a simplification of the results in which the colored area represents the range between the 16-th and the 84-th percentiles. The thick lines represent the mean fit functions ($f_1,f_2,f_3$) obtained for each algorithm, calculated as the mean of the parameters for each run.}
    \label{fig:results}
\end{figure}

As already mentioned, measuring the performance of a subroutine is a complex task, since it is not possible to directly access the cost function of the problem. Thus, we propose a metric for measuring the performance of the sampling strategy that consist of counting the number of times we iterate the subroutine for measuring the full feasible partial solution space $\mathcal{F}$. This way, this metric provides an estimation of the number of runs we would need to ensure we can generate the optimal solutions to the full problem. 

For testing the quantum annealing subroutine, we performed its classical simulation. Nevertheless, as already discussed before (Section \ref{sec:algSamplingSubroutine}), this would give an estimation for the performance of the algorithm when implemented in a real quantum device. Since the original annealing subroutine was designed for a fixed amount of runs, we have slightly modified the algorithm for this work. Instead of performing a sweep over different values of $\alpha$, we fixed its value so that the ground state corresponds to configurations with total weights in the middle of the range of allowed values, i.e. configurations with total weights of $C/2$. The parameters we have employed for the simulation are $\beta=\min(w)/5$, $\alpha=C\beta$, $T=10^{-14}$, $\gamma=10$, $n_\text{T}=500$, $\lVert H_0\rVert=10^n$ and $\gamma=2$, with $\hbar=6.58\cdot10^{-16}$.

A consideration one has to take into account when analysing these results is that the quantum annealing and the hybrid approach both have a set of hyperparameters that affects their performance. As in classical heuristic optimizers, selecting a good set of hyperparameters can increase the performance of the algorithms. However, obtaining such parameters is usually a difficult task, which involves running the algorithm for a large training set of instances and obtaining an intuition about how each of the parameters affects the result. In this work, we have not perform such analysis due to the computational cost of simulating quantum annealing processes. However, we are confident that a better selection of hyperparameters could greatly enhance the performance of these approaches. 

For the implementation of the hybrid strategy, we have employed the same parameters in the annealing part. The switching point has been defined with the following heuristic. First, we obtained a linear fit for the function $f(x)=a\cdot x$, where $f(x)$ is the number of different feasible partial solutions obtained until the iteration $x$. With this, we consider that the random walk is still efficient while the expected probability of obtaining a new result is higher than $25\%$ (i.e. $a>1/4$) or while RMSD to respect to the linear fit is lower than 2. On top of this, we force the algorithm to perform a random walk for the first 100 iterations. It is important to highlight that these rules and parameter values have been manually selected, based on obtained results. However, these selection of hyperparameters could be further optimized.

We performed a benchmark over 18 instances of 10 and 12 packages and different package weight distributions, using the same instances as in \cite{mikelgda2022BPPEXTRA}. We implemented the algorithm proposals from section \ref{sec:algSamplingSubroutine}, running the algorithm twice per instance (to make the best use of the computational resources we had available). Additionally, due to the computational cost of simulating the quantum annealing for larger systems, we let both the simulation of the quantum annealing and the hybrid sampling run for a maximum amount of time. The results we have obtained for the benchmark are shown in Fig. \ref{fig:results}. The key observation regarding the performance of the algorithms is the point in the $x$ axis at which each line reaches the point $y=1$. This value gives us the information about how faster does any of the three algorithms completes the sampling compared to the worst algorithm for each instance. As expected, the random sampling gives the worst possible performance in almost every case we have tested. On the other hand, the random walk strategy is the fastest to fully sample $\mathcal{F}$. However, the random walk strategy clearly shows stagnation when sampling the last of the feasible partial solutions. We argue that the random walk can outperform the rest of the algorithms for small problems, but for larger ones, the probability of measuring new feasible partial solutions would drastically decrease. The proposed hybrid sampling shows an improvement over the random walk in 4 out of the 9 runs in which every algorithm sampled the full space $\mathcal{F}$.

For gaining intuition on the performance of all the algorithms tested, we fit the values obtained to functions that mimics their expected general behaviour. As we expect the random sampling strategy to be the worst possible one, we fit the results to a exponential function $f_1(x,a) = 1-2^{-ax}+x2^{-a}$ for $a>0$, where here $x$ is the fraction of iterations until the sample is completed. We have selected this exponential function to show the exponential decay of the probability of measuring new results, and to have two fixed points at $f_1(0,a)=0$ and $f_1(1,a)=1$. As we expect the random walk to show a similar behaviour, we employed a similar function, but with an extra parameter to show a slightly different staggering compared to the random sampling $f_2(x,a,b) = 1-2^{-ax}+x2^{-b}$. For the annealing subroutine we see that the fraction of feasible solutions sampled grows close to linearly. Thus, we select a simple linear fit $f_3(x,a) = ax$. As we expect the hybrid approach to maintain the linear behaviour of the annealing, we employed same fit function as for the random walk. Indeed, we see that the asymptotic behaviour of the hybrid strategy improves the one of the random walk, albeit slightly. Although the results we obtained in this work can not qualify as a proof, we are confident that this hybrid approach would vastly improve the performance of the random walk for problems with more packages.

\section{Conclusion}\label{sec:conclusion}

In summary, we have compared the quantum annealing algorithm for sampling partial solutions of the 1dBPP against other classical algorithms. We have proposed two classical algorithms for the problem of sampling feasible partial solutions, a random strategy and a heuristic based on the random walk. We show that the asymptotic behaviour of both classical strategies stagnates when trying to completely sample the subset of feasible partial solutions to the 1dBPP. To overcome this problem, we have proposed an extra hybrid sampling strategy where we use the quantum annealing approach when efficiency of the classical strategy decreases. This strategy not only improves the results of the subroutine in terms of number of iterations, but also uses the quantum resources less times. In order to validate these hypothesis, we have performed a benchmark over 18 instances comparing every algorithm proposed in this paper. The results we obtained show that the hybrid approach for the sampling problem obtains an advantage over the rest of the approaches. Furthermore, this paves the way for employing the quantum algorithms in a more efficient manner, only when classical algorithms can not further improve their results. This gives a new perspective, opening the possibility to obtain new hybrid algorithms or improve the already existing ones. 

An important remark about the simulations we have performed is that, due to the limited time and computational resources available, we have not perform any optimization of the hyperparameters. Even so, the results still show a clear advantage for the quantum and the hybrid quantum-classical approaches. This also gives us a intuition about the expected performance of the algorithm for larger problems. Optimizing the parameters for the quantum annealing would increase the fidelity of the process, and thus improving the results we have already obtained. 

As follow-up work for the research done in this paper, one can extend the comparison of the sampling subroutine to other quantum algorithms. This would enrich the variety of subroutines one can employ to solve the full problem. An analysis on how to combine different strategies could help improving the full hybrid algorithm beyond the results we already obtained. 

A natural question that arises is whether the full hybrid algorithm gives any advantage over other classical optimizers. However, before answering this, the subroutine building the full solutions to the problem should be optimized. Optimizing both of the subroutines separately should provide the best performance for the full algorithm. However, finding a fair metric to compare classical, quantum and hybrid algorithms is still an open question. Furthermore, giving an answer for the underlying question of the quantum advantage is a challenging research topic itself \cite{Boixo2018Supremacy, Arute2019google, WeiPan202261qubitDAQCEXTRA}.


\section*{Acknowledgements}

The research leading to this paper has received funding from the QUANTEK project (ELKARTEK program from the Basque Government, expedient no. KK-2021/00070). Additionaly, IO, EVR and EO acknowledges support from the Spanish CDTI through Misiones Ciencia e Innovación Program (CUCO) under Grant MIG-20211005. MS acknowledges support from Spanish Ram\'on y Cajal Grant RYC-2020-030503-I and the project grant PID2021-125823NA-I00 funded by MCIN/AEI/10.13039/501100011033 and by ``ERDF A way of making Europe" and ``ERDF Invest in your Future", the projects QMiCS (820505) and OpenSuperQ (820363) of the EU Flagship on Quantum Technologies, from the EU FET Open project Quromorphic (828826) and EPIQUS (899368). MGdA acknowledges support from the UPV/EHU and TECNALIA 2021 PIF contract call.

\bibliographystyle{IEEEtran}    
\bibliography{biblio.bib, biblioEXTRA.bib} 

\end{document}